\documentclass[11pt]{article}
\usepackage{cite}
\usepackage{epsfig}
\usepackage{amsmath}
\usepackage{amssymb}
\newcommand{\ben}{\begin{eqnarray}}
\newcommand{\een}{\end{eqnarray}}
\newcommand{\nnu}{\nonumber\\}
\newcommand{\bef}{\begin{figure}[htb]\centering}
\newcommand{\eef}{\end{figure}}

\begin{document}

\title{\bf General positivity bounds for spin observables \\
in single particle inclusive production}


\author{Zhong-Bo Kang$^{1}$\footnote{zkang@bnl.gov} ~and
            Jacques Soffer$^{2}$\footnote{jacques.soffer@gmail.com}
\\[0.3cm]
{\normalsize \it $^1$RIKEN BNL Research Center, Brookhaven National Laboratory, Upton, NY 11973, USA}
\\[0.1cm]
{\normalsize \it $^2$Physics Department, Barton Hall, Temple University, Philadelphia, PA 19122, USA}}
\maketitle

\begin{abstract}
\noindent
Positivity constraints, derived initially assuming parity conservation, 
for the inclusive reaction of the type $A(\mbox{spin 1/2})+B(\mbox{spin 1/2})\to C+X$,
where the spins of both initial spin-$1/2$ particles can be in any possible directions and no polarization is observed in the final state, are generalized to the case of parity violation. By means of a systematic method, we obtain non-trivial bounds involving all the spin observables of the reaction and we discuss some relevant physics processes. Particularly we discover a non-trivial
positivity constraint for the processes, $pp\to W^{\pm}/Z^0+X$ or $pp\to \ell^{\pm}+X$ where $\ell^{\pm}$ decayed from $W^{\pm}/Z^0$, which could be checked at the ongoing longitudinal spin program at RHIC.

\end{abstract}

\section{Introduction}

Positivity constraints have been widely studied in spin physics to pin down 
the smallest allowed domain for spin observables. This powerful tool has a 
broad range of applications for exclusive reactions as well as for inclusive 
reactions. It can be used to test the consistency of a given set of available 
measurements and also the validity of specific dynamical assumptions 
in theoretical models. Different methods can be used to establish these 
constraints and many interesting cases have been presented in a recent 
review article \cite{Artru:2008cp}. 

In the present paper, we will focus on the single particle inclusive production
in polarized hadronic collisions, $A(\mbox{spin 1/2})+B(\mbox{spin 1/2})\to C+X$, 
where only initial spins are observed and no polarization for the final state particle 
$C$ is measured. If parity is conserved, this reaction is fully described in terms 
of {\it eight} independent spin observables. The positivity constraints for these
{\it parity-conserving} observables have been derived in Ref.~\cite{Soffer:2003qj}.
However, if parity is not conserved, there are twice as many spin observables and
the positivity constraints become much more involved. Nevertheless, such a case has 
been partially studied in Ref.~\cite{Kang:2010fu}. The derived positivity constraint
has been further used to put non-trivial bounds on several Sivers functions \cite{Kang:2010fu}, entering the theoretical description of single transverse spin asymmetry for various processes \cite{Kang:2009bp, Kang:2009sm, Anselmino:2008sga, Anselmino:2009st, Gamberg:2010tj, Kang:2011hk}. However, the increasing complexity for parity-violation case requires a systematic method to obtain all the positivity constraints. This will be the aim of this paper, i.e., deriving all the positivity constraints for
the most general cases - including both {\it parity-conserving} and {\it parity-violating} 
spin observables. 

The longitudinal $W^{\pm}$ program at Relativistic Heavy Ion Collider (RHIC) at Brookhaven
National Laboratory is currently under successful running, aiming to pin down the polarized
antiquark distribution in the proton \cite{Bunce:2000uv, spinplan}. The reason that $W^{\pm}$ boson production could provide
unique and clean access to the individual antiquark polarizations is due to 
the maximal violation of parity in the elementary $Wq\bar{q}'$ vertex \cite{Bourrely:1993dd,Bourrely:1994sc}. Because of the {\it parity-violation} nature of this process,
$pp\to W^{\pm}+X$ or $pp\to \ell^{\pm}+X$ where $\ell^{\pm}$ decayed from $W^{\pm}$, our
newly derived positivity constraints could have interesting and nontrivial implications for
the spin asymmetries measured in these processes. Spin asymmetries for the process $pp\to Z^0+X$ can be also measured, although
the production rate is lower than for $W^{\pm}$ production.

The remainder of this paper is organized as follows: in the next section we review the derivation of positivity constraints for the parity-conservation case, to introduce the notation and also to update the method which could be easily used for the parity-violation case. In section 3, we will derive all the general positivity constraints, which involve both parity-conserving and parity-violating spin observables, and are classified according to different degrees, linear, quadratic, cubic and quartic. We then give one phenomenological example of our derived positivity bound, when applying to the inclusive $W^{\pm}/Z^0$ production in longitudinal $pp$ scatterings. We summarize our results in section 5.

\section{Positivity constraints for parity-conserving processes}

For single particle inclusive production: $A(\mbox{spin 1/2})+B(\mbox{spin 1/2})\to C+X$ with spin vectors $P_a$ and $P_b$ for initial particles $A$ and $B$, respectively, 
the spin-dependent cross section $\sigma(P_a, P_b)$ can be defined through the cross section matrix $M$ and the spin density matrix $\rho$
\ben
\sigma\left(P_a, P_b\right)={\rm Tr}\left(M\rho\right),
\een
where $\rho=\rho_a\otimes \rho_b$ is the spin density matrix with $\rho_i=(I_2+P_i\cdot \vec{\sigma})/2, i=a, b$. Here $\vec{\sigma}=(\sigma_x, \sigma_y, \sigma_z)$ stands for the three $2\times 2$ Pauli matrices and $I_2$ is the $2\times 2$ unit matrix, thus $\rho$ being the direct product of $\rho_a$ and $\rho_b$ will be $4\times 4$ matrix. We will now study the parametrization of the $4\times 4$ cross section matrix $M$. We first review the existing study for the parity-conserving case. The purpose is twofold: first to set up notation; and second, to update the method for deriving the positivity constraints, which could also be used in the parity-violation case. 

For the {\it parity-conserving} process, $M$ could be parametrized in the following way
\ben
M&=&\sigma_0\left[I_4+A_{aN}\sigma_{ay}\otimes I_2+A_{bN}I_2\otimes \sigma_{by}
+A_{NN}\sigma_{ay}\otimes \sigma_{by}+A_{LL}\sigma_{az}\otimes \sigma_{bz}
\right.
\nnu
&&\left.
+A_{SS}\sigma_{ax}\otimes \sigma_{bx}+A_{LS}\sigma_{az}\otimes\sigma_{bx}
+A_{SL}\sigma_{ax}\otimes\sigma_{bz}\right].
\label{Mpc}
\een
Here $I_4$ is the $4\times 4$ unit matrix and $\sigma_0$ stands for the spin-averaged cross
section. In other words, for a parity-conserving process, there are {\it eight} independent
spin-dependent observables \cite{Artru:2008cp, Soffer:2003qj, Goldstein:1975ci}: the unpolarized
cross section $\sigma_0$, {\it two} single transverse spin asymmetries $A_{aN}$ and $A_{bN}$, and {\it five} double spin asymmetries $A_{NN}$, $A_{LL}$, $A_{SS}$, $A_{LS}$, and $A_{SL}$. 
Here the subscript $L$, $N$, $S$ represents the unit vectors along the spin directions of initial particles $A$ and $B$. Specifically in the center-of-mass system of $A$ and $B$, $L$, $N$, $S$ are along the incident momentum, along the normal to the scattering plane which
contains $A$, $B$ and $C$, and along $N\times L$, respectively. The expression in Eq.~(\ref{Mpc}) is fully justified, since we have explicitly
\ben
\sigma\left(P_a, P_b\right)&=&\sigma_0[1+A_{aN}P_{ay}+A_{bN}P_{by}
+A_{NN}P_{ay}P_{by}+A_{LL}P_{az}P_{bz}
\nnu
&&
+A_{SS}P_{ax}P_{bx}+A_{LS}P_{az}P_{bx}
+A_{SL}P_{ax}P_{bz}].
\een

The positivity constraints for the parity-conserving process have been derived in \cite{Soffer:2003qj}. The crucial point is that the cross section matrix $M$ is a Hermitian and positive matrix. The {\it necessary} and {\it sufficient} condition for a Hermitian matrix to be positive is that {\it all its eigenvalues are positive}. It is important to emphasize here that the eigenvalues of a matrix are independent of the basis where it is written. In other words, no matter what basis one chooses to express the cross section matrix $M$, one should obtain the {\it same} set of eigenvalues, from which one obtains the {\it same} set of positivity bounds. Since this is the {\it necessary} and {\it sufficient} condition, the bounds derived from this should be the strongest constraints for the spin observables. Using this fact about the eigenvalues, we thus could choose any convenient basis such that the matrix has a simple form and thus the eigenvalues could be easily derived from there. For example, the original positivity bounds are derived by choosing the transverse basis where $\sigma_y$ is diagonal, in which the matrix elements $M_{ij}$ of the cross section matrix $M$ are given by
\ben
\begin{array}{l}
M_{11}=\left(1+A_{NN}\right)+\left(A_{aN}+A_{bN}\right),
\\
M_{22}=\left(1-A_{NN}\right)+\left(A_{aN}-A_{bN}\right),
\\
M_{33}=\left(1-A_{NN}\right)-\left(A_{aN}-A_{bN}\right),
\\
M_{44}=\left(1+A_{NN}\right)-\left(A_{aN}+A_{bN}\right),
\\
M_{14}=M_{41}^*=A_{LL}-A_{SS}-i\left(A_{SL}+A_{LS}\right),
\\
M_{23}=M_{32}^*=A_{LL}+A_{SS}-i\left(A_{SL}-A_{LS}\right),
\\
M_{12}=M_{21}=M_{13}=M_{31}=M_{24}=M_{42}=M_{34}=M_{43}=0,
\end{array}
\label{mpcsigmay}
\een
i.e., half of the matrix elements vanish in this basis. The eigenvalues $\lambda_i$ can be easily obtained and are given by
\ben
\lambda_{1, 2}&=&1+A_{NN}\pm \sqrt{(A_{aN}+ A_{bN})^2+(A_{LL}- A_{SS})^2+(A_{LS}+A_{SL})^2}
\\
\lambda_{3, 4}&=&1-A_{NN}\pm \sqrt{(A_{aN}- A_{bN})^2+(A_{LL}+ A_{SS})^2+(A_{LS}-A_{SL})^2}
\een
Then from all the eigenvalues $\lambda_i\geq 0$, we have the following strongest positivity constraints
\ben
(1\pm A_{NN})^2\geq (A_{aN}\pm A_{bN})^2+(A_{LL}\mp A_{SS})^2+(A_{LS}\pm A_{SL})^2,
\label{pctot}
\een
which are exactly the same as those derived in \cite{Soffer:2003qj}. 

It is nice to be able to derive the eigenvalues of the cross section matrix and thus to obtain the strongest positivity bounds directly. However, sometimes the eigenvalues turn out to be very difficult to find, which is exactly the situation we face for the parity-violation case in the next section. 
In such situations, one could derive a complete set of {\it necessary} constraints which as a whole forms the {\it sufficient} condition for positivity. This is the strategy of J.J. Sylvester, so-called {\it Sylvester's criterion}, which states that the {\it necessary} and {\it sufficient} condition for a Hermitian matrix to be semi-positive is that {\it all its principal minors have to be non-negative}. This is equivalent to say for the $4\times 4$ matrix $M$ that: 
\begin{itemize}
\item All the diagonal matrix elements $M_{ii}\geq 0$;
\item The elements satisfy: $M_{ii}M_{jj}\geq |M_{ij}|^2$;
\item The determinant of any $3\times 3$ matrix formed by removing from $M$ its $i^{\rm th}$ row and $i^{\rm th}$ column are non-negative, where $i=1,2,3$, or 4;
\item The determinant of the matrix $M$ itself is non-negative.
\end{itemize}
These conditions will enable us to derive all the general positivity constraints, which are of different degree, linear, quadratic, cubic and quartic, respectively, and as a whole they form the {\it necessary} and {\it sufficient} condition for the positivity of $M$.

Let's use Sylvester's criterion to revisit the positivity bounds for parity-conserving case. First let's work again in the transverse basis.  With the expression of $M$ in Eq.~(\ref{mpcsigmay}) in hand, from $M_{ii}\geq 0$ we immediately derive
\ben
1\pm A_{NN}>\left|A_{aN}\pm A_{bN}\right|.
\label{pcdiag}
\een
On the other hand, from $M_{ii}M_{jj}\geq |M_{ij}|^2$, we further derive
\ben
(1\pm A_{NN})^2\geq (A_{aN}\pm A_{bN})^2+(A_{LL}\mp A_{SS})^2+(A_{SL}\pm A_{LS})^2,
\label{pctott}
\een 
It is also easy to find that going to the even higher order principal minors does not give any further constraints. Thus we have derived the equivalent results from a slightly different method. 

We would also like to make some comments on the positivity bound in Eq.~(\ref{pcdiag}). Although this is a weaker bound and could be deduced from Eq.~(\ref{pctot}), its derivation is much simpler in the sense that one only deals with the diagonal matrix elements without looking for the eigenvalues which need extra (and difficult) mathematical work. On the other hand, these kinds of linear bounds are also very useful in the phenomenological studies because of the limited accessibility and accuracy of the spin asymmetries in the experimental measurements.

It is also important to realize that Sylvester's criterion holds true in any basis one chooses to express the cross section matrix $M$.  Thus choosing a different basis, one could derive a different set of positivity bounds. Even though these different sets are equivalent to each other according to Sylvester's criterion, one might obtain some interesting positivity bounds by choosing a convenient basis, which could be difficult to find directly in another basis.

For example, if we choose the helicity basis where $\sigma_z$ is diagonal, the cross section matrix $M$ becomes
\ben
\begin{array}{l}
M_{11}=M_{44}=1+A_{LL} 
\\
M_{22}=M_{33}=1-A_{LL}
\\
M_{14}=M_{41}=A_{SS}-A_{NN}
\\
M_{23}=M_{32}=A_{SS}+A_{NN}
\\
M_{12}=M_{21}^*=A_{LS}-i A_{bN} 
\\
M_{13}=M_{31}^*=A_{SL}+i A_{aN}
\\
M_{24}=M_{42}^*=-A_{SL}-i A_{aN} 
\\
M_{34}=M_{43}^*=-A_{LS}-i A_{bN}
\end{array}
\een
Now $M_{ii}\geq 0$ leads to the trivial bounds $1\pm A_{LL}\geq 0$. However, using $M_{ii}M_{jj}\geq |M_{ij}|^2$ leads to some new positivity bounds:
\ben
1-A_{LL}^2 &\geq& A_{LS}^2+A_{bN}^2,
\label{sigza}
\\
1-A_{LL}^2 &\geq& A_{SL}^2+A_{aN}^2,
\label{sigzb}
\\
(1\pm A_{LL})^2&\geq& (A_{SS}\mp A_{NN})^2,
\label{sigzc}
\een
which are not very easy to derive in the transverse basis. These are actually weaker bounds compared with the strongest bounds in Eq.~(\ref{pctot}). It requires some extra work to derive them in the transverse basis: if one uses the following formula in Eq.~(\ref{pctot})
\ben
\sqrt{a_1^2+a_2^2+\cdots +a_n^2}\geq \left| a_1+a_2+\cdots + a_n\right|,
\een
one could derive all the inequalities (\ref{sigza}) - (\ref{sigzc}). One could continue to study the higher order principal minors and thus to obtain the cubic and quartic constraints. Then together with the linear and quadratic ones, they form the complete set of positivity bounds which should be equivalent to the bounds in Eq.~(\ref{pctot}) according to Sylvester's criterion.

\section{All general positivity constraints}

We now study the general positivity constraints, which involve also the {\it parity-violating} asymmetries where one will have {\it sixteen} independent spin-dependent observables \cite{Kang:2010fu}. Besides those in the parity-conserving processes, one has {\it four} additional
single spin asymmetries $A_{aL}$, $A_{bL}$, $A_{aS}$ and $A_{bS}$, and {\it four} additional 
double spin asymmetries $A_{LN}$, $A_{NL}$, $A_{NS}$ and $A_{SN}$. In this case, the more general cross section matrix $M$ can be parametrized as
\ben
M&=&\sigma_0\left[I_4+A_{aN}\sigma_{ay}\otimes I_2+A_{bN}I_2\otimes \sigma_{by}
+A_{NN}\sigma_{ay}\otimes \sigma_{by}+A_{LL}\sigma_{az}\otimes \sigma_{bz}
\right.
\nnu
&&\left.
+A_{SS}\sigma_{ax}\otimes \sigma_{bx}+A_{LS}\sigma_{az}\otimes\sigma_{bx}
+A_{SL}\sigma_{ax}\otimes\sigma_{bz}\right]
\nnu
&&
+\sigma_0\left[
A_{aL}\sigma_{az}\otimes I_2+A_{bL}I_2\otimes \sigma_{bz}
+A_{aS}\sigma_{ax}\otimes I_2+A_{bS}I_2\otimes \sigma_{bx}
\right.
\nnu
&&
\left.
+A_{LN}\sigma_{az}\otimes \sigma_{by}+A_{NL}\sigma_{ay}\otimes \sigma_{bz}
+A_{NS}\sigma_{ay}\otimes \sigma_{bx}+A_{SN}\sigma_{ax}\otimes \sigma_{by}\right],
\label{Mpv}
\een
which is fully justified since one has
\ben
\sigma\left(P_a, P_b\right)&=&\sigma_0[1+A_{aN}P_{ay}+A_{bN}P_{by}
+A_{NN}P_{ay}P_{by}+A_{LL}P_{az}P_{bz}
\nnu
&&+A_{SS}P_{ax}P_{ax}+A_{LS}P_{az}P_{bx}
+A_{SL}P_{ax}P_{bz}]
\nnu
&&
+\sigma_0[A_{aL}P_{az}+A_{bL}P_{bz}
+A_{aS}P_{ax}+A_{bS}P_{bx}
\nnu
&&
+A_{LN}P_{az}P_{by}+A_{NL}P_{ay}P_{bz}
+A_{NS}P_{ay}P_{bx}+A_{SN}P_{ax}P_{by}
].
\een
Again to study the positivity constraints, we need to express the cross section matrix $M$ in a basis and then study the {\it necessary} and {\it sufficient} conditions for $M$ to be positive. Because of the complexity of the matrix $M$, it becomes difficult to obtain the eigenvalues. Instead we will use Sylvester's criterion.

First, we will still work in the transverse basis where $\sigma_y$ is diagonal, in which the matrix elements of $M$ are given by
\ben
\begin{array}{l}
M_{11}=\left(1+A_{NN}\right)+\left(A_{aN}+A_{bN}\right)
\\
M_{22}=\left(1-A_{NN}\right)+\left(A_{aN}-A_{bN}\right)
\\
M_{33}=\left(1-A_{NN}\right)-\left(A_{aN}-A_{bN}\right)
\\
M_{44}=\left(1+A_{NN}\right)-\left(A_{aN}+A_{bN}\right)
\\
M_{14}=M_{41}^*=A_{LL}-A_{SS}-i\left(A_{SL}+A_{LS}\right)
\\
M_{23}=M_{32}^*=A_{LL}+A_{SS}-i\left(A_{SL}-A_{LS}\right)
\\
M_{12}=M_{21}^*=A_{bL}+A_{NL}-i \left(A_{bS}+A_{NS}\right)
\\
M_{13}=M_{31}^*=A_{aL}+A_{LN}-i \left(A_{aS}+A_{SN}\right)
\\
M_{24}=M_{42}^*=A_{aL}-A_{LN}-i \left(A_{aS}-A_{SN}\right)
\\
M_{34}=M_{43}^*=A_{bL}-A_{NL}-i \left(A_{bS}-A_{NS}\right)
\end{array}
\label{pvsigmay}
\een
Start with the linear positivity bounds. From $M_{ii}>0$, we immediately obtain
\ben
1\pm A_{NN}>\left|A_{aN}\pm A_{bN}\right|.
\label{pvdiag}
\een
The quadratic bounds are derived from $M_{ii}M_{jj}\geq |M_{ij}|^2$. Especially for $\{i,j\}=\{1,4\}$, and $\{2,3\}$, one obtains
\ben
(1\pm A_{NN})^2\geq (A_{aN}\pm A_{bN})^2+(A_{LL}\mp A_{SS})^2+(A_{SL}\pm A_{LS})^2.
\label{pvtot}
\een
Both Eqs.~(\ref{pvdiag}) and (\ref{pvtot}) are exactly the same as those in Eqs.~(\ref{pcdiag}) and (\ref{pctott}), which are derived for the parity-conserving case. Since all the asymmetries involved in these inequalities are parity-conserving ones, it is not surprising that these positivity bounds are preserved. For the case when $\{i,j\}=\{1,2\}, \{1,3\}, \{2,4\}$, and $\{3,4\}$, we obtain four extra positivity bounds
\ben
(1\pm A_{aN})^2&\geq& (A_{bN}\pm A_{NN})^2+(A_{bL}\pm A_{NL})^2+(A_{bS}\pm A_{NS})^2,
\\
(1\pm A_{bN})^2&\geq& (A_{aN}\pm A_{NN})^2+(A_{aL}\pm A_{LN})^2+(A_{aS}\pm A_{SN})^2,
\een
which involve both parity-conserving and parity-violating asymmetries.

The cubic bounds are derived from the determinants of all the $3\times 3$ matrix formed by removing from $M$ its $i^{\rm th}$ row and column, which has to be non-negative. If we remove the $4^{\rm th}$ row and column, i.e., keep the row 1, 2, 3 and column 1, 2, 3, this newly formed matrix denoted as $M_{123}$ is given by
\begin{small}
\ben
M_{123}=\left(
\begin{array}{lll}
A_{aN}+A_{bN}+A_{NN}+1 
& A_{bL}+A_{NL}-i (A_{bS}+A_{NS})
& A_{aL}+A_{LN}-i (A_{aS}+A_{SN})
\\ 
A_{bL}+A_{NL}+i (A_{bS}+A_{NS})
&A_{aN}-A_{bN}-A_{NN}+1
& A_{LL}+A_{SS}-i(A_{SL}-A_{LS})
\\
A_{aL}+A_{LN}+i (A_{aS}+A_{SN})
& A_{LL}+A_{SS}+i (A_{SL}-A_{LS})
& -A_{aN}+A_{bN}-A_{NN}+1
\end{array}
\right)
\een
\end{small}
Then the determinant $|M_{123}|\geq 0$ leads to the following inequality
\ben
|M_{123}|&=&(1+A_{NN}+A_{aN}+A_{bN})[(1-A_{NN})^2-(A_{aN}-A_{bN})^2-(A_{LL}+A_{SS})^2-(A_{SL}-A_{LS})^2]
\nnu
&\ &
-(A_{aN}-A_{bN}-A_{NN}+1)[(A_{bL}+A_{NL})^2+(A_{bS}+A_{NS})^2]
\nnu
&\ &
-(-A_{aN}+A_{bN}-A_{NN}+1)[(A_{aL}+A_{LN})^2+(A_{aS}+A_{SN})^2]
\nnu
&\ &
+2(A_{LL}+A_{SS})[(A_{bL}+A_{NL})(A_{aL}+A_{LN})+(A_{bS}+A_{NS})(A_{aS}+A_{SN})]
\nnu
&\ &
+2(A_{LS}-A_{SL})[(A_{bS}+A_{NS})(A_{aL}+A_{LN})-(A_{bL}+A_{NL})(A_{aS}+A_{SN})]
\geq 0
\een
Likewise, from $|M_{124}|, |M_{134}|, |M_{234}| \geq 0$, we have
\ben
|M_{124}|&=&(1-A_{NN}+A_{aN}-A_{bN})[(1+A_{NN})^2-(A_{aN}+A_{bN})^2-(A_{LL}-A_{SS})^2-(A_{SL}+A_{LS})^2]
\nnu
&\ &
-(1+A_{NN}-A_{aN}-A_{bN})[(A_{bL}+A_{NL})^2+(A_{bS}+A_{NS})^2]
\nnu
&\ &
-(1+A_{NN}+A_{aN}+A_{bN})[(A_{aL}-A_{LN})^2+(A_{aS}-A_{SN})^2]
\nnu
&\ &
+2(A_{aL}-A_{LN})[(A_{bL}+A_{NL})(A_{LL}-A_{SS})+(A_{bS}+A_{NS})(A_{SL}+A_{LS})]
\nnu
&\ &
+2(A_{aS}-A_{SN})[(A_{bL}+A_{NL})(A_{SL}+A_{LS})-(A_{bS}+A_{NS})(A_{LL}-A_{SS})]
\geq 0
\een
\ben
|M_{134}|&=&(1-A_{NN}-A_{aN}+A_{bN})[(1+A_{NN})^2-(A_{aN}+A_{bN})^2-(A_{LL}-A_{SS})^2-(A_{SL}+A_{LS})^2]
\nnu
&\ &
-(1+A_{NN}-A_{aN}-A_{bN})[(A_{aL}+A_{LN})^2+(A_{aS}+A_{SN})^2]
\nnu
&\ &
-(1+A_{NN}+A_{aN}+A_{bN})[(A_{bL}-A_{NL})^2+(A_{bS}-A_{NS})^2]
\nnu
&\ &
+2(A_{bL}-A_{NL})[(A_{aL}+A_{LN})(A_{LL}-A_{SS})+(A_{aS}+A_{SN})(A_{SL}+A_{LS})]
\nnu
&\ &
+2(A_{bS}-A_{NS})[(A_{aL}+A_{LN})(A_{SL}+A_{LS})-(A_{aS}+A_{SN})(A_{LL}-A_{SS})]
\geq 0
\een
\ben
|M_{234}|&=&(1+A_{NN}-A_{aN}-A_{bN})[(1-A_{NN})^2-(A_{aN}-A_{bN})^2-(A_{LL}+A_{SS})^2-(A_{SL}-A_{LS})^2]
\nnu
&\ &
-(1-A_{NN}-A_{aN}+A_{bN})[(A_{aL}-A_{LN})^2+(A_{aS}-A_{SN})^2]
\nnu
&\ &
-(1-A_{NN}+A_{aN}-A_{bN})[(A_{bL}+A_{NL})^2+(A_{bS}+A_{NS})^2]
\nnu
&\ &
+2(A_{bL}-A_{NL})[(A_{LL}+A_{SS})(A_{aL}-A_{LN})+(A_{SL}-A_{LS})(A_{aS}-A_{SN})]
\nnu
&\ &
+2(A_{bS}-A_{NS})[(A_{LL}+A_{SS})(A_{aS}-A_{SN})-(A_{SL}-A_{LS})(A_{aL}-A_{LN})]
\geq 0
\een
Finally the quartic bounds are given by the determinant of $M$ itself $|M|\geq 0$. With the matrix elements given in Eq.~(\ref{pvsigmay}), it is easy to write down the determinant of $M$, which we do not write out explicitly here. The linear, quadratic, cubic and quartic bounds form the complete set of positivity bounds according to Sylvester's criterion. We have found that all the bounds derived from the parity-conserving case are preserved in this more general case.

As we have emphasized in the previous section, we could also derive a different set of positivity bounds by choosing a different basis. These new bounds might also be useful and interesting for the phenomenological studies.  Let us study the positivity bounds by choosing the helicity basis, in which the explicit form of $M$ is given by
\ben
\begin{array}{l}
M_{11}=\left(1+A_{LL}\right)+\left(A_{aL}+A_{bL}\right)
\\
M_{22}=\left(1-A_{LL}\right)+\left(A_{aL}-A_{bL}\right)
\\
M_{33}=\left(1-A_{LL}\right)-\left(A_{aL}-A_{bL}\right)
\\
M_{44}=\left(1+A_{LL}\right)-\left(A_{aL}+A_{bL}\right)
\\
M_{14}=M_{41}^*=A_{SS}-A_{NN}-i\left(A_{NS}+A_{SN}\right)
\\
M_{23}=M_{32}^*=A_{SS}+A_{NN}-i\left(A_{NS}-A_{SN}\right)
\\
M_{12}=M_{21}^*=A_{bS}+A_{LS}-i \left(A_{bN}+A_{LN}\right)
\\
M_{13}=M_{31}^*=A_{aS}+A_{SL}-i \left(A_{aN}+A_{NL}\right)
\\
M_{24}=M_{42}^*=A_{aS}-A_{SL}-i \left(A_{aN}-A_{NL}\right)
\\
M_{34}=M_{43}^*=A_{bS}-A_{LS}-i \left(A_{bN}-A_{LN}\right)
\end{array}
\label{pvsigmaz}
\een
Now from $M_{ii}\geq 0$, we have
\ben
1\pm A_{LL}>|A_{aL}\pm A_{bL}|.
\label{pvlong}
\een
This is a very interesting positivity bound. Particularly it involves both parity-conserving ($A_{LL}$) and parity-violating ($A_L$) asymmetries. Since this bound involves only the longitudinal asymmetries, it might be very useful and relevant to the ongoing longitudinal $W^{\pm}/Z^0$ program at RHIC.

Then from $M_{ii}M_{jj}\geq |M_{ij}|^2$, one could derive the following quadratic bounds
\ben
(1\pm A_{LL})^2&\geq& (A_{aL}\pm A_{bL})^2+(A_{SS}\mp A_{NN})^2+(A_{NS}\pm A_{SN})^2,
\label{pvspe}
\\
(1\pm A_{aL})^2&\geq& (A_{bL}\pm A_{LL})^2+(A_{bS}\pm A_{LS})^2+(A_{bN}\pm A_{LN})^2,
\\
(1\pm A_{bL})^2&\geq& (A_{aL}\pm A_{LL})^2+(A_{aS}\pm A_{SL})^2+(A_{aN}\pm A_{NL})^2.
\een
These bounds are stronger than the linear ones. For example, specifically from the bounds in Eq.~(\ref{pvspe}), one could deduce those in Eq.~(\ref{pvlong}). One could continue to study the bounds in even higher order, and the procedure is straightforward. We decide to stop here, and turn to discuss some phenomenological applications.

\section{Phenomenological example: $W^{\pm}/Z^0$ production at RHIC}
In the last two sections, we have derived quite a few positivity constraints which involve both parity-conserving and parity-violating spin asymmetries. They could have broad applications in testing the consistency of the experimental measurements, or studying the validity of the theoretical models. However, due to the experimental limited accessibility and accuracy, only a few could be reachable in the near future. In this section, we study one such example: the positivity bound (\ref{pvlong}) in the $W^{\pm}/Z^0$ or $\ell^{\pm}$ production in $pp$ collisions, $pp\to W^{\pm}/Z^0+X$ or $pp\to \ell^{\pm}+X$ where $\ell^{\pm}$ decayed from $W^{\pm}/Z^0$. It is important to realize that for identical initial particles scattering, one has
\ben
A_{aL}(y)=A_{bL}(-y),
\een
where $y$ is the rapidity of the final-state particle. Thus Eq.~(\ref{pvlong}) becomes
\ben
1\pm A_{LL}(y)>|A_{L}(y)\pm A_{L}(-y)|.
\label{longbound}
\een
These bounds have a very simple form, and should be very interesting to test them in RHIC experiments.

To have an idea, let us check whether these bounds are satisfied for $W^{\pm}$ or $Z^0$ production in longitudinal $pp$ collisions, $\vec{p}\vec{p}\to W^{\pm}/Z^0+X$. In perturbative QCD formalism, at leading-order and restricting to only up and down quarks, one has the following simple expressions for the single spin asymmetries \cite{Bourrely:1993dd,Bourrely:1994sc,Gehrmann:1997ez,deFlorian:2010aa,vonArx:2011fz}
\ben
A^{W^+}_{L} (y)& = & \frac{-\Delta u(x_a) \bar d(x_b) +
\Delta \bar d(x_a) u(x_b)}{u(x_a) 
\bar d(x_b) + \bar d(x_a) u(x_b)}\; , 
\\
A^{W^-}_{L} (y)& = & \frac{- \Delta d(x_a) \bar u(x_b) +
\Delta \bar u(x_a) d(x_b)}{d(x_a) 
\bar u(x_b) + \bar u(x_a) d(x_b)}\; ,  
\\
A^{Z^{0}}_{L} (y) & = & \frac{\sum_q (2 v_q a_q) 
\left[-\Delta q(x_a) \bar q(x_b) +
\Delta \bar q(x_a) q(x_b) \right] }
{ \sum_q (v_q^2 + a_q^2) 
\left[  q(x_a) \bar q(x_b) + 
\bar q(x_a)  q(x_b) \right] }\; ,
\een
and for the double spin asymmetries
\ben
A^{W^+}_{LL} (y)& = & - \frac{\Delta u(x_a) \Delta \bar d(x_b) +
\Delta \bar d(x_a) \Delta u(x_b)}{u(x_a) 
\bar d(x_b) + \bar d(x_a) u(x_b)}\; , 
\\
A^{W^-}_{LL} (y)& = & - \frac{\Delta d(x_a) \Delta \bar u(x_b) +
\Delta \bar u(x_a) \Delta d(x_b)}{d(x_a) 
\bar u(x_b) + \bar u(x_a) d(x_b)}\; ,
\\
A^{Z^{0}}_{LL} (y) & = & - \frac{\sum_q (v_q^2 + a_q^2) 
\left[ \Delta q(x_a) \Delta \bar q(x_b) + 
\Delta \bar q(x_a) \Delta q(x_b) \right] }
{ \sum_q (v_q^2 + a_q^2) 
\left[  q(x_a) \bar q(x_b) + 
\bar q(x_a)  q(x_b) \right] }\; ,
\een
where $\Delta q(x)$ and $q(x)$ are the helicity distribution and unpolarized parton distribution function, respectively. $v_q$ and $a_q$ are the vector and axial couplings of the $Z^0$ boson to the quark. $x_{a,b}$ are the parton momentum fractions given by
\ben
x_a=m_Q/\sqrt{s} \,e^{y},
\qquad
x_b=m_Q/\sqrt{s} \,e^{-y},
\een
with $m_Q, y$ the mass and rapidity of the $W$ (or $Z$) boson and $\sqrt{s}$ the center-of-mass energy.
\bef
\psfig{file=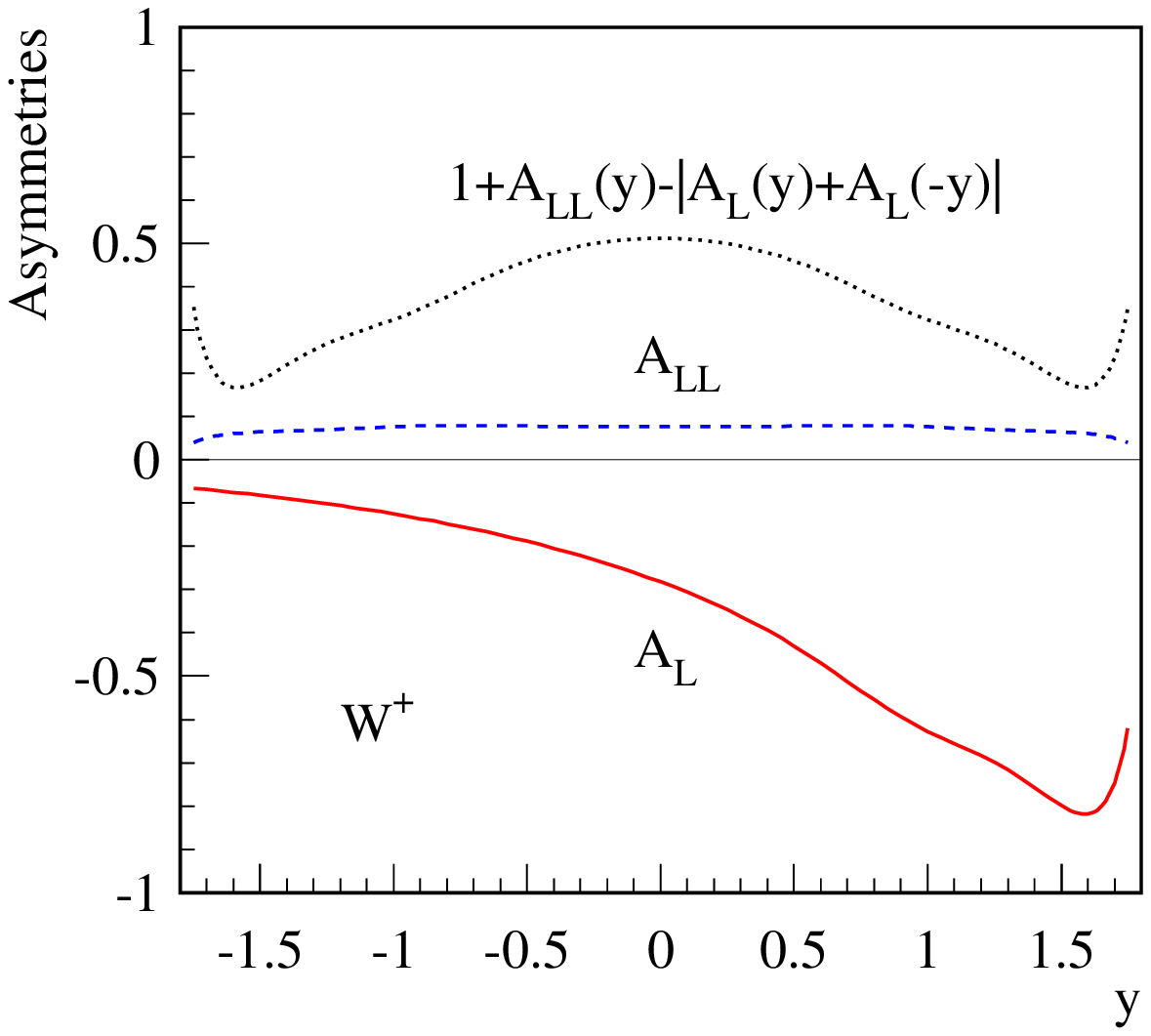, width=2.6in}
\hskip 0.4in
\psfig{file=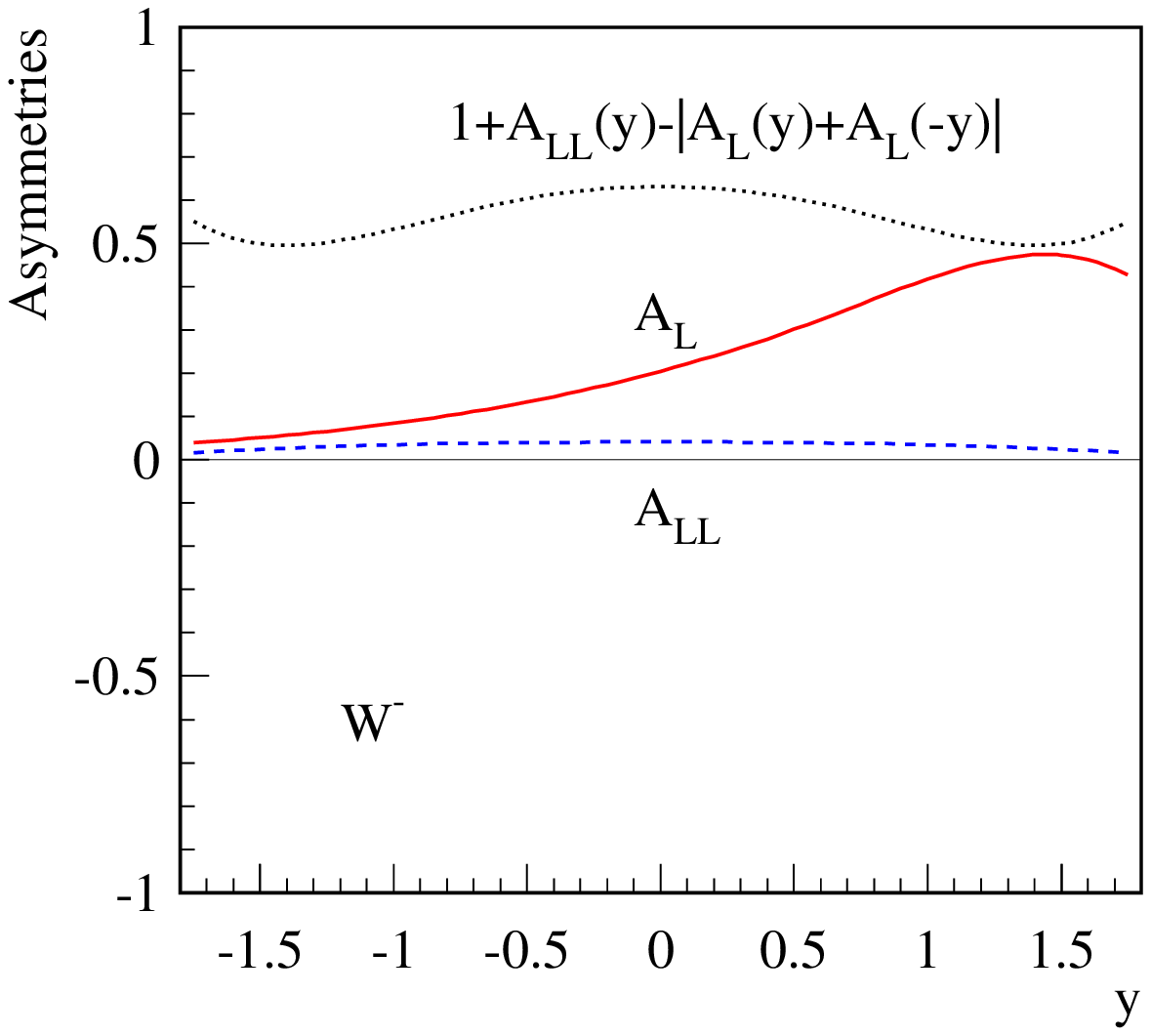, width=2.6in}
\caption{\small Longitudinal asymmetries are plotted as a function of rapidity $y$ of the $W$ boson in $\vec{p}\vec{p}$ collisions: $W^+$ (left) and $W^-$ (right). The solid curves are the single longitudinal spin asymmetry $A_L$, the dashed curves are the double longitudinal spin asymmetry $A_{LL}$, and the dotted curves are the combination of $1+A_{LL}(y)-|A_L(y)+A_L(-y)|$.}
\label{asymmetry}
\eef
\bef
\psfig{file=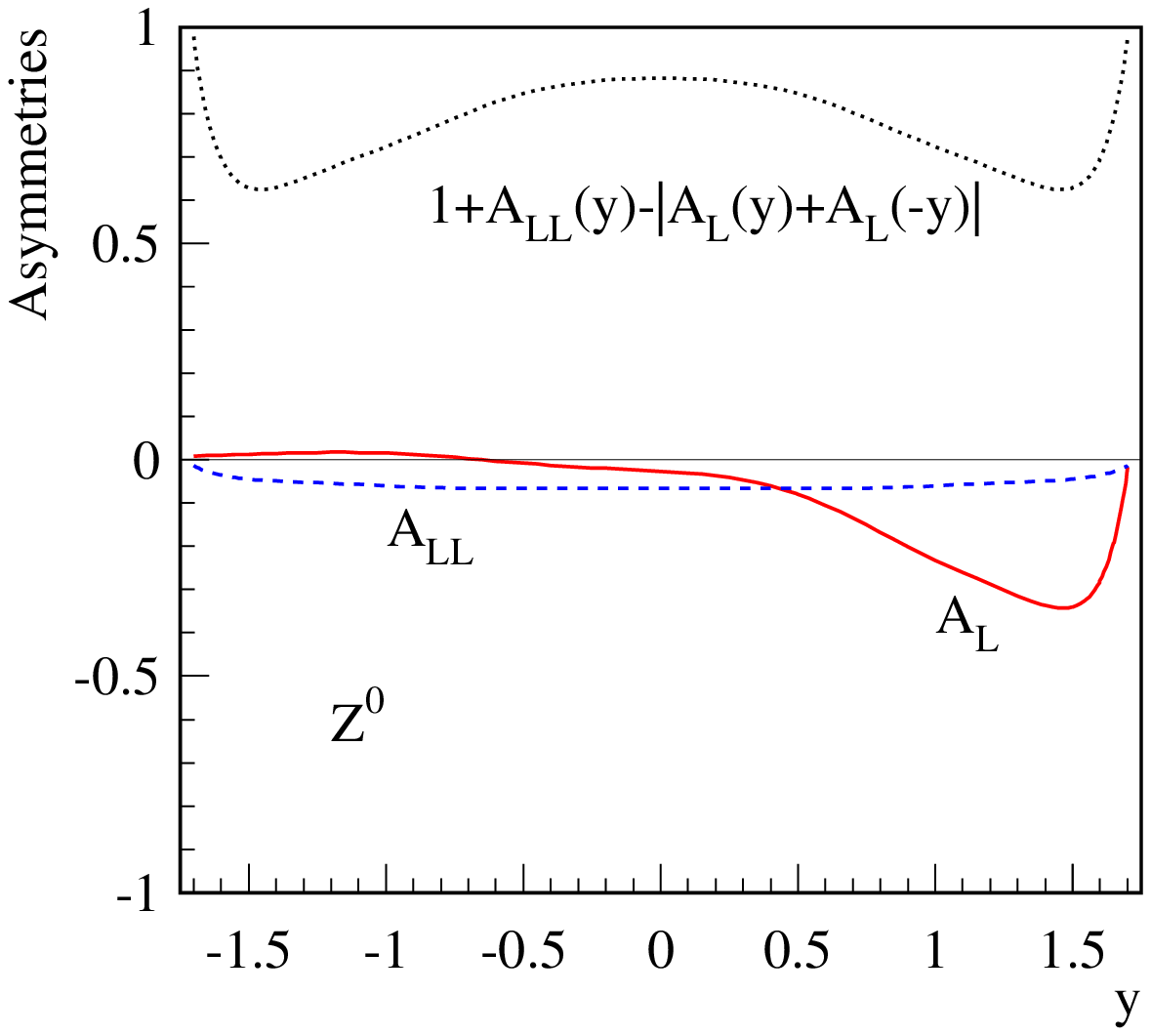, width=2.6in}
\caption{\small Same as Fig.~\ref{asymmetry}, but for $Z^0$ boson.}
\label{asymmetryZ}
\eef

To estimate these asymmetries numerically, we choose BBS2001 polarized and unpolarized parton distribution functions based on a statistical approach \cite{Bourrely:2001du}. At $\sqrt{s}=500$ GeV, our calculations plotted as a function of rapidity $y$ are shown in Fig.~\ref{asymmetry} for $W^+$ (left) and $W^-$ (right), and in Fig.~\ref{asymmetryZ} for $Z^0$ boson, respectively. The solid curves are the single longitudinal spin asymmetry $A_L$, the dashed curves are the double longitudinal spin asymmetry $A_{LL}$, while the dotted curves are the following combination 
\ben
1+A_{LL}(y)-|A_L(y)+A_L(-y)|,
\een
which must be positive according to the positivity bounds in Eq.~(\ref{longbound}). As we can see from both plots that even though $|A_L(y)+A_L(-y)|$ could become quite sizable, it is still smaller than $1+A_{LL}(y)$, thus the bound is satisfied in this leading order calculation for both $W^{\pm}$ and $Z^0$.  The other bound $1-A_{LL}(y)-|A_L(y)-A_L(-y)|$ could become even more sizable and is also satisfied within BBS2001 parameterization.

We have also checked other popular parametrizations of polarized parton distribution functions, to see whether they satisfy our bounds. Particularly we have checked GRSV2000 \cite{Gluck:2000dy}, AAC2008 \cite{Hirai:2008aj}, DSSV \cite{deFlorian:2008mr, deFlorian:2009vb}, and LSS2010 \cite{Leader:2010rb}. For the unpolarized parton distribution functions, we use exactly the same set as the one when the global fitting was performed. That is, we use GRV98 \cite{Gluck:1998xa} for both GRSV2000 and AAC2008, while MRST2002 \cite{Martin:2002aw} for both DSSV and LSS2010. It turns out that both GRSV2000 and AAC2008 satisfy our bounds, while both DSSV and LSS2010 could have violation at large rapidity $|y|$. The violations are shown in Fig.~\ref{violation}, in which we plot $1+A_{LL}(y)-|A_L(y)+A_L(-y)|$ as a function of rapidity $y$ for $W^+$, $W^-$, and $Z^0$. Since for identical incoming hadrons $A_{LL}(-y)=A_{LL}(y)$, the combination $1+A_{LL}(y)-|A_L(y)+A_L(-y)|$ is symmetric under $y\leftrightarrow -y$ and we thus only plot for positive $y$. We immediately find that when rapidity becomes large $y\gtrsim 1.5$ where $x_a\gtrsim 0.7$ and $x_b\lesssim 0.04$ (or vice verse for $y\lesssim -1.5$), the combination $1+A_{LL}(y)-|A_L(y)+A_L(-y)|$ could become negative for both DSSV and LSS2010, and for all $W^{\pm}/Z^0$ bosons. Since our bounds are very general, coming from the positivity conditions of the cross section matrix, they should always be satisfied. Thus our newly derived bounds, though very simple, immediately put nontrivial constraints on the parametrizations of the polarized parton distributions, for both DSSV and LSS2010. 

\bef
\psfig{file=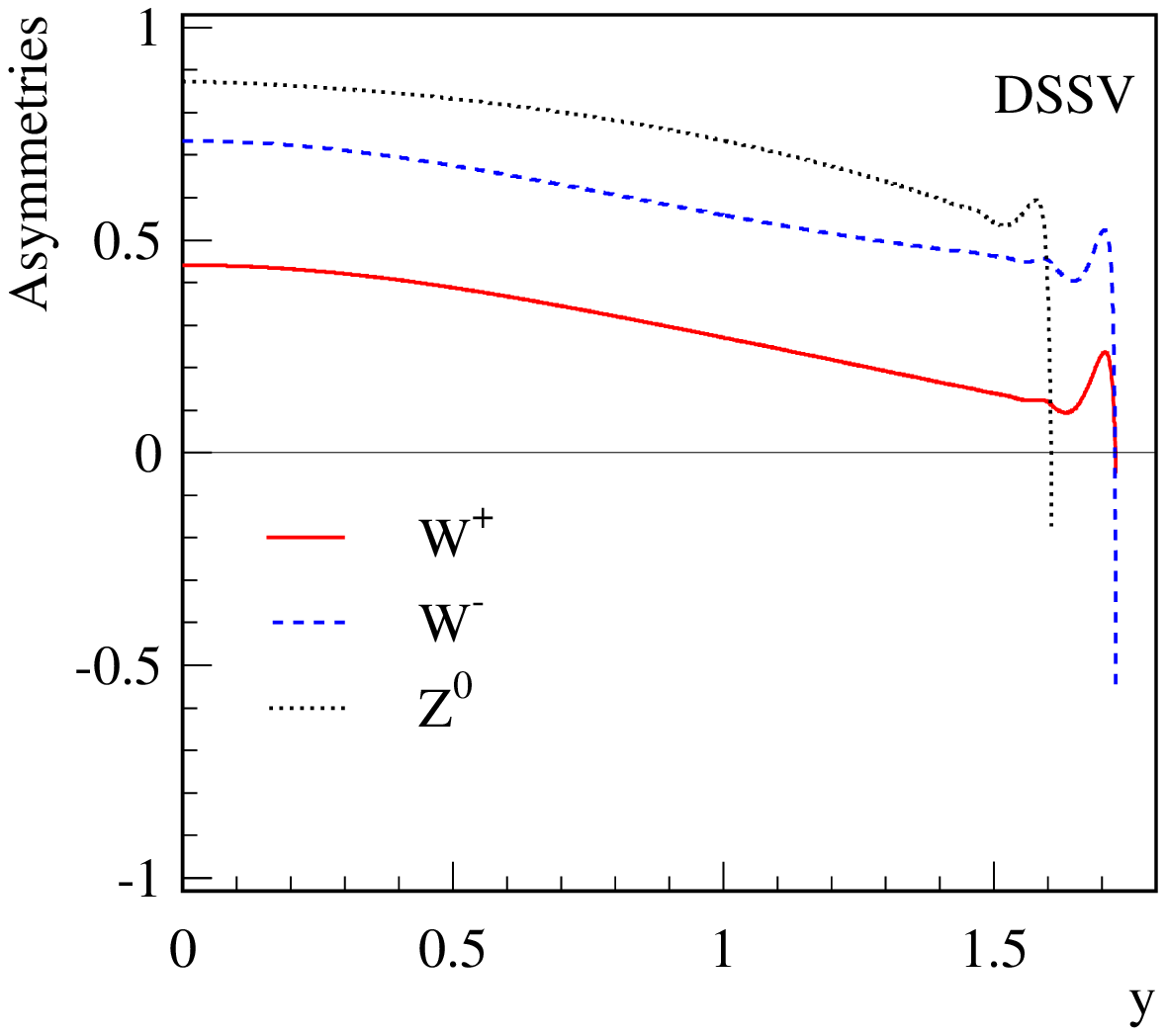, width=2.6in}
\hskip 0.4in
\psfig{file=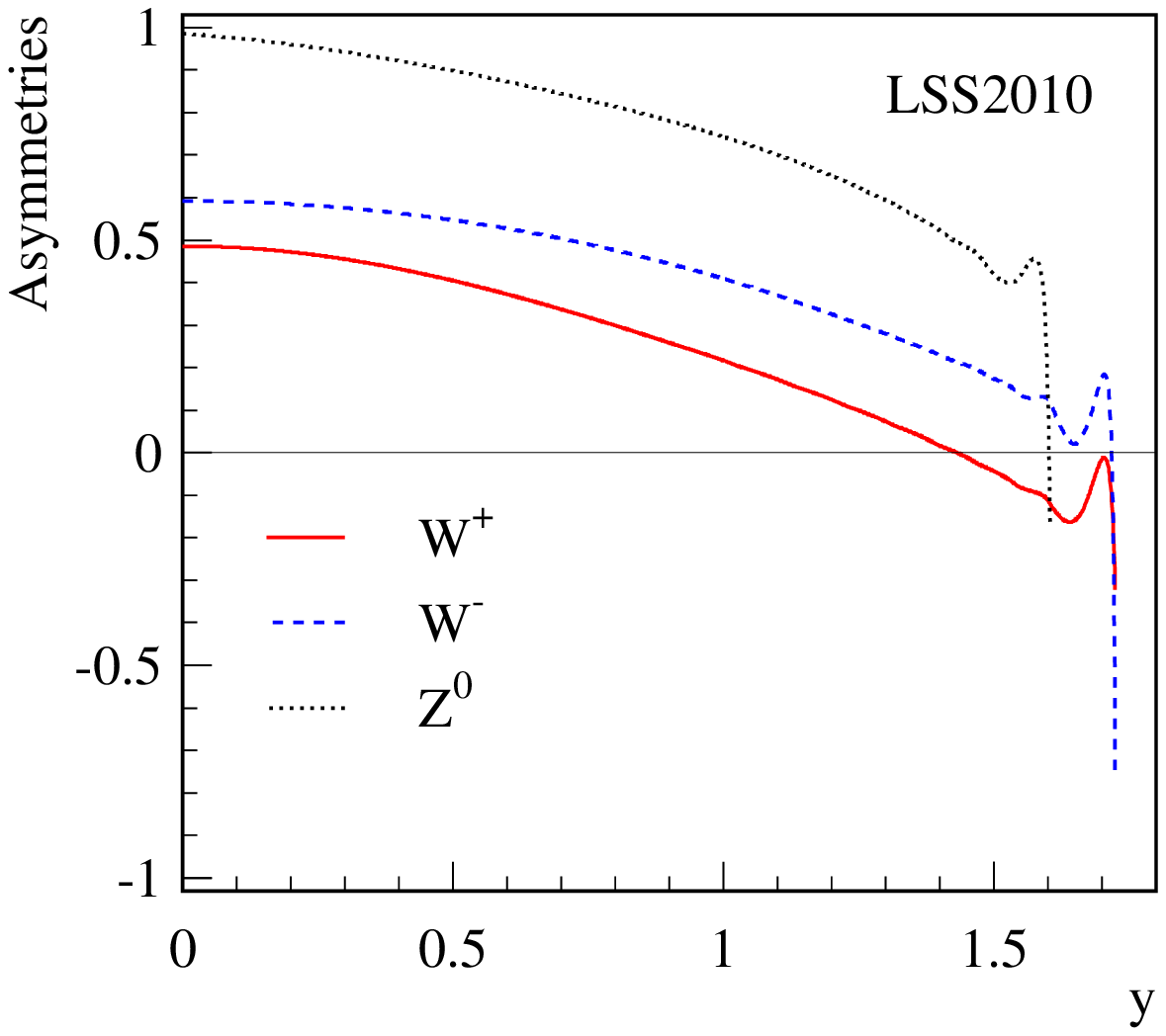, width=2.6in}
\caption{Asymmetry $1+A_{LL}(y)-|A_L(y)+A_L(-y)|$ are plotted as a function of rapidity $y$ for the parametrization of DSSV \cite{deFlorian:2008mr, deFlorian:2009vb} (left) and LSS2010 \cite{Leader:2010rb} (right). The solid, dashed and dotted lines are for $W^+$, $W^-$, and $Z^0$, respectively.}
\label{violation}
\eef
According to next-to-leading order calculations in \cite{deFlorian:2010aa, vonArx:2011fz}, the QCD radiative corrections for the asymmetries are small. We thus expect our findings and conclusions for the specific parametrizations of polarized parton distribution functions will not alter in higher-order QCD calculations, for both $W^{\pm}/Z^0$, and for the leptons decayed from them. It will be very interesting to check our bounds in the forward or backward regions (when $|y|$ becomes large) in the experiments. 

We notice that there are already published data for single spin asymmetry $A_L$ at mid-rapidity $y=0$ from both STAR \cite{Aggarwal:2010vc} and PHENIX \cite {Adare:2010xa}, for the leptons decayed from $W^{\pm}/Z^0$ bosons.
\ben
\mbox{STAR: } &&
A_L^{e^+}=-0.27\pm 0.10, \qquad 
A_L^{e^-}=0.14\pm 0.19
\\
\mbox{PHENIX: } &&
A_L^{e^+}=-0.86^{+0.30}_{-0.14}, \qquad 
A_L^{e^-}=0.88^{+0.12}_{-0.71}
\een
On the other hand, at $y=0$ our bounds becomes 
\ben
1+A_{LL}(0) \geq 2|A_L(0)|.
\label{midbound}
\een
At the same time $A_{LL}(0)$ is very small (a few percents) for the current parametrization of polarized parton distribution functions. If one takes $A_{LL}(0)\approx 0$ in Eq.~(\ref{midbound}), we obtain 
\ben
|A_L(0)| \leq \frac{1}{2}.
\een
Comparing with both STAR and PHENIX data, we immediately find that STAR data is consistent with our bounds, while PHENIX data (central value) is certainly out of the bound. Of course, so far the data has very large uncertainty at the moment. Nonetheless, it will be important to check all our bounds in the future experiments. We look forward to the future experimental measurements to test these bounds.

\section{Summary}
We derive all the general positivity constraints for the spin observables in the single particle inclusive production, $A(\mbox{spin 1/2})+B(\mbox{spin 1/2})\to C+X$, where the spins of both initial spin-$1/2$ particles can be in any possible directions and no polarization is observed in the final state. By means of a systematic method, we generalize the previous positivity constraints derived for the parity-conserving processes to the most general processes including also the parity-violating ones. We find that the positivity constraints involving only the parity-conserving asymmetries are preserved in the parity-violating case. 

With the help of Sylvester's criterion, we derive all the general positivity constraints, which are of different degree, linear, quadratic, cubic and quartic, respectively. These constraints form a complete set of {\it necessary} and {\it sufficient} condition for positivity. As a special example, we discover a very interesting non-trivial bound for the parity-conserving and parity-violating longitudinal asymmetries $A_{LL}$ and $A_L$. This could be relevant to the processes, $pp\to W^{\pm}/Z^0+X$ or $pp\to \ell^{\pm}+X$ where $\ell^{\pm}$ decayed from $W^{\pm}/Z^0$ in longitudinal $pp$ scatterings, which are currently under active investigation at RHIC.

The positivity constraints derived here could have broad applications in testing the consistency of the experimental measurements, or studying the validity of the theoretical models. We look forward to the future experimental data to test these positivity bounds.

Before closing, let us mention another possible application of our results in polarized deep inelastic scattering at high energies where weak
interactions contributions, both neutral and charged current processes are taken into account, as well as the parity violating polarized nucleon structure functions. The explicit expressions of the charged lepton asymmetry $A_{eL}(y)$ and the proton asymmetry
$A_{pL}(y)$ can be obtained from Ref.~\cite{Anselmino:1993tc} and our bounds can be used to put new constraints on the helicity distributions. These parity-violating asymmetries are also expected to be modified by effects beyond the Standard Model, {\it e.g.} scalar and vector 
leptoquarks \cite{Taxil:1999pf}, and our bounds could be used to put further constraints on existing models.

\vskip 0.5cm
\noindent
{\bf Acknowledgments:} 
We thank Andreas Metz and Werner Vogelsang for discussions.
We are grateful to RIKEN, Brookhaven National Laboratory, and the U.S.~Department
of Energy (Contract No.~DE-AC02-98CH10886) for supporting this work. 



\begin{thebibliography}{99}

\bibitem{Artru:2008cp}
  X.~Artru, M.~Elchikh, J.~M.~Richard, J.~Soffer and O.~V.~Teryaev,
  Phys.\ Rept.\  {\bf 470}, 1 (2009)
  [arXiv:0802.0164 [hep-ph]].

\bibitem{Soffer:2003qj}
  J.~Soffer,
  Phys.\ Rev.\ Lett.\  {\bf 91}, 092005 (2003)
  [arXiv:hep-ph/0305222].

\bibitem{Kang:2010fu}
  Z.~B.~Kang and J.~Soffer,
  Phys.\ Lett.\  B {\bf 695}, 275 (2011)
  [arXiv:1003.4913 [hep-ph]].
  
\bibitem{Kang:2009bp}
  Z.~B.~Kang and J.~W.~Qiu,
  Phys.\ Rev.\ Lett.\  {\bf 103}, 172001 (2009)
  [arXiv:0903.3629 [hep-ph]].
  
\bibitem{Kang:2009sm}
  Z.~B.~Kang and J.~W.~Qiu,
  Phys.\ Rev.\  D {\bf 81}, 054020 (2010)
  [arXiv:0912.1319 [hep-ph]].

\bibitem{Anselmino:2008sga}
  M.~Anselmino {\it et al.},
  Eur.\ Phys.\ J.\  A {\bf 39}, 89 (2009)
  [arXiv:0805.2677 [hep-ph]].

\bibitem{Anselmino:2009st}
  M.~Anselmino, M.~Boglione, U.~D'Alesio, S.~Melis, F.~Murgia and A.~Prokudin,
  Phys.\ Rev.\  D {\bf 79}, 054010 (2009)
  [arXiv:0901.3078 [hep-ph]].

\bibitem{Gamberg:2010tj}
  L.~Gamberg and Z.~B.~Kang,
  Phys.\ Lett.\  B {\bf 696}, 109 (2011)
  [arXiv:1009.1936 [hep-ph]].

\bibitem{Kang:2011hk}
  Z.~B.~Kang, J.~W.~Qiu, W.~Vogelsang and F.~Yuan,
  arXiv:1103.1591 [hep-ph].

\bibitem{Bunce:2000uv}
  G.~Bunce, N.~Saito, J.~Soffer and W.~Vogelsang,
  Ann.\ Rev.\ Nucl.\ Part.\ Sci.\  {\bf 50}, 525 (2000)
  [arXiv:hep-ph/0007218].

\bibitem{spinplan} G.\ Bunce {\it et al.}, ``Plans for the RHIC Spin Physics
  Program'', June 2008, see: {\tt spin.riken.bnl.gov/rsc}.

\bibitem{Bourrely:1993dd}
  C.~Bourrely and J.~Soffer,
  Phys.\ Lett.\  B {\bf 314}, 132 (1993).

\bibitem{Bourrely:1994sc}
  C.~Bourrely and J.~Soffer,
  Nucl.\ Phys.\  B {\bf 423}, 329 (1994)
  [arXiv:hep-ph/9405250].

\bibitem{Goldstein:1975ci}
  G.~R.~Goldstein and J.~F.~Owens,
  Nucl.\ Phys.\  B {\bf 103}, 145 (1976).

\bibitem{Gehrmann:1997ez}
  T.~Gehrmann,
  Nucl.\ Phys.\  B {\bf 534}, 21 (1998)
  [arXiv:hep-ph/9710508].

\bibitem{deFlorian:2010aa}
  D.~de Florian and W.~Vogelsang,
  Phys.\ Rev.\  D {\bf 81}, 094020 (2010)
  [arXiv:1003.4533 [hep-ph]].

\bibitem{vonArx:2011fz}
  C.~von Arx and T.~Gehrmann,
  arXiv:1103.1465 [hep-ph].

\bibitem{Bourrely:2001du}
  C.~Bourrely, J.~Soffer and F.~Buccella,
  Eur.\ Phys.\ J.\  C {\bf 23}, 487 (2002)
  [arXiv:hep-ph/0109160].

\bibitem{Gluck:2000dy}
  M.~Gluck, E.~Reya, M.~Stratmann and W.~Vogelsang,
  Phys.\ Rev.\  D {\bf 63}, 094005 (2001)
  [arXiv:hep-ph/0011215].

\bibitem{Hirai:2008aj}
  M.~Hirai and S.~Kumano  [Asymmetry Analysis Collaboration],
  Nucl.\ Phys.\  B {\bf 813}, 106 (2009)
  [arXiv:0808.0413 [hep-ph]].


\bibitem{deFlorian:2008mr}
  D.~de Florian, R.~Sassot, M.~Stratmann and W.~Vogelsang,
  Phys.\ Rev.\ Lett.\  {\bf 101}, 072001 (2008)
  [arXiv:0804.0422 [hep-ph]].
  
\bibitem{deFlorian:2009vb}
  D.~de Florian, R.~Sassot, M.~Stratmann and W.~Vogelsang,
  Phys.\ Rev.\  D {\bf 80}, 034030 (2009)
  [arXiv:0904.3821 [hep-ph]].

\bibitem{Leader:2010rb}
  E.~Leader, A.~V.~Sidorov and D.~B.~Stamenov,
  Phys.\ Rev.\  D {\bf 82}, 114018 (2010)
  [arXiv:1010.0574 [hep-ph]].

\bibitem{Gluck:1998xa}
  M.~Gluck, E.~Reya and A.~Vogt,
  Eur.\ Phys.\ J.\  C {\bf 5}, 461 (1998)
  [arXiv:hep-ph/9806404].

\bibitem{Martin:2002aw}
  A.~D.~Martin, R.~G.~Roberts, W.~J.~Stirling and R.~S.~Thorne,
  Eur.\ Phys.\ J.\  C {\bf 28}, 455 (2003)
  [arXiv:hep-ph/0211080].


\bibitem{Aggarwal:2010vc}
  M.~M.~Aggarwal {\it et al.}  [STAR Collaboration],
  Phys.\ Rev.\ Lett.\  {\bf 106}, 062002 (2011)
  [arXiv:1009.0326 [hep-ex]].

\bibitem{Adare:2010xa}
  A.~Adare {\it et al.}  [PHENIX Collaboration],
  Phys.\ Rev.\ Lett.\  {\bf 106}, 062001 (2011)
  [arXiv:1009.0505 [hep-ex]].
  
\bibitem{Anselmino:1993tc}
  M.~Anselmino, P.~Gambino and J.~Kalinowski,
  Z.\ Phys.\  C {\bf 64}, 267 (1994)
  [arXiv:hep-ph/9401264].

\bibitem{Taxil:1999pf}
  P.~Taxil, E.~Tugcu and J.~M.~Virey,
  Eur.\ Phys.\ J.\  C {\bf 14}, 165 (2000)
  [arXiv:hep-ph/9912272].


\end{thebibliography}
\end{document}